\documentclass[conference]{IEEEtran}
\usepackage{graphicx,
	psfrag,
	epsfig,
	epstopdf,
    enumerate,
	amsthm,
    amsmath,
	amssymb,
	url,
	algorithm,
	algorithmic,
	balance,
	enumerate,
	color,
    url,
	setspace,
	tikz,
	pgfplots,
    geometry,
    placeins,
    cleveref,
    multirow,
    array,
}
\usepackage{amsmath}
\usepackage[nospace,noadjust]{cite}
\newgeometry{left=1.6cm,right=1.6cm,bottom=1.6cm, top=1.6cm}

\newcommand{\ignore}[1]{}

\begin{document}
\title{Machine Learning Towards Enabling Spectrum-as-a-Service Dynamic Sharing}
\author{
\IEEEauthorblockN{Abdallah Moubayed\IEEEauthorrefmark{1}, Tanveer Ahmed\IEEEauthorrefmark{2}, Anwar Haque\IEEEauthorrefmark{1}, and Abdallah Shami\IEEEauthorrefmark{1}}
		
\IEEEauthorblockA{\IEEEauthorrefmark{1} Western University, London, Ontario, Canada, 
	emails: \{amoubaye, ahaque32, abdallah.shami\}@uwo.ca
}
\IEEEauthorblockA{\IEEEauthorrefmark{2} Nordicity, Ottawa, Ontario, Canada, 
	e-mail: tahmed@nordicity.com 
}
}
\maketitle

\begin{abstract}
The growth in wireless broadband users, devices, and novel applications has led to a significant increase in the demand for new radio frequency spectrum. This is expected to grow even further given the projection that the global traffic per year will reach  4.8 zettabytes by 2022. Moreover, it is projected that the number of Internet users will reach 4.8 billion and the number of connected devices will be close 28.5 billion devices. However, due to the spectrum being mostly allocated and divided, providing more spectrum to expand existing services or offer new ones has become more challenging. To address this, spectrum sharing has been proposed as a potential solution to improve spectrum utilization efficiency. Adopting effective and efficient spectrum sharing mechanisms is in itself a challenging task given the multitude of levels and techniques that can be integrated to enable it. To that end, this paper provides an overview of the different spectrum sharing levels and techniques that have been proposed in the literature. Moreover, it discusses the potential of adopting dynamic sharing mechanisms by offering Spectrum-as-a-Service architecture. Furthermore, it describes the potential role of machine learning models in facilitating the automated and efficient dynamic sharing of the spectrum and offering Spectrum-as-a-Service.    
\end{abstract}


\section{Introduction}\label{Intro}
\indent The growth in wireless broadband users, devices, and novel applications has led to a significant increase in the demand for new radio frequency spectrum. This is expected to grow even further given the projection that the global traffic per year will reach  4.8 zettabytes by 2022 \cite{Cisco_data_growth}. Moreover, it is projected that the number of Internet users will reach 4.8 billion and the number of connected devices will be close 28.5 billion devices \cite{Cisco_data_growth}. However, due to the spectrum being mostly allocated and divided \cite{us_freq_allocation_2019}, providing more spectrum to expand existing services or offer new ones has become more challenging. 
Nonetheless, recent studies have shown that the issue is not the lack of spectrum, but rather spectrum access \cite{spectrum_inefficiency1}. This means that the spectrum’s capacity is not being exploited to its full extent \cite{spectrum_inefficiency1}. This is mainly due to the exclusive use licensing model adopted by spectrum regulators globally in which incumbent operators under-utilize the spectrum they hold the license for \cite{spectrum_inefficiency2}. Moreover, service providers (SPs) are searching for creative ways to meet the growth in data services’ demand rate for new bandwidth-intensive services and applications such as video and music streaming while simultaneously improving the average gain per user \cite{Cisco_data_growth}. This is especially critical for fifth generation (5G) networks where the service requirements are extremely stringent \cite{5G_requirements}.\\
\indent One potential solution to the spectrum under-utilization is spectrum sharing. Spectrum sharing allows for the radio resources to be re-used by multiple users/technologies/applications to improve the spectrum efficiency and increase the throughput without needing new resources. The benefit of spectrum sharing was illustrated by a study conducted in Europe that showed that the spectrum needed to support 5G networks can be reduced from 76 GHz (if spectrum is used exclusively) to 19 GHz if spectrum sharing is enabled \cite{spectrum_sharing1}. Accordingly, several spectrum regulators have allow multiple technologies to access increasing number of shared frequency bands under novel access right frameworks \cite{spectrum_sharing2,spectrum_sharing3,spectrum_sharing4}. Moreover, it was shown that spectrum sharing allows SPs to reduce their capital and operating expenses due to the cost-sharing process involved \cite{spectrum_sharing_cost_benefit}.\\
\indent The spectrum sharing process can be implemented at different levels and using different techniques. For example, spectrum sharing can be done in a static or dynamic manner (either based on previously agreed on ratios or opportunistically). Also, the spectrum access can be done based on frequency sharing, time sharing, or even space sharing. Moreover, sharing can be implemented at the technology level either within the same technology or across multiple technologies \cite{spectrum_sharing3,spectrum_sharing4}. The different levels at which spectrum sharing can be enabled and the multitude of techniques that can be used to achieve it introduces a myriad of challenges \cite{spectrum_sharing_challenges}. More specifically,managing the spectrum sharing process in a dynamic and efficient manner while maintaining different regulatory and performance constraints is a challenging task \cite{spectrum_sharing_challenges}.\\
\indent To that end, machine learning (ML) can play a major role in facilitating the dynamic sharing of the spectrum efficiently. Using the huge data generated by the different technologies and spectrum sensing techniques, ML algorithms can extract insightful information about the characteristics and behavior of systems \cite{ML_networking}. As a result, these algorithms can make more informed decisions on how to allocate the spectrum \cite{ML_spectrum_sharing1,ML_spectrum_sharing2,ML_spectrum_sharing3}. The potential of ML is further illustrated by the substantial growth in filled ML patents filled in the US at a compound annual growth rate of 34\% between 2013 to 2017 \cite{ML_statistic1}.\\ 
\indent This paper provides an overview of the different spectrum sharing levels and techniques that have been proposed in the literature. Moreover, it discusses the potential of adopting dynamic sharing mechanisms by offering Spectrum-as-a-Service architecture. Furthermore, it describes the potential role of machine learning models in facilitating the automated and efficient dynamic sharing of the spectrum. \\
\indent The paper is structured as follows: Section \ref{spectrum_sharing} provides an overview of the different levels and techniques proposed in the literature for spectrum sharing. Section \ref{ML} briefly introduces the concept of ML and presents some of the previous works focusing on spectrum sharing. Moreover, it describes the future opportunities that ML techniques offer as key enablers of dynamic spectrum sharing through a Spectrum-as-a-Service architecture. Finally, Section \ref{conc} concludes the paper.
\section{Spectrum Sharing Techniques} \label{spectrum_sharing}
\begin{figure}[!h]
	\centering
	\includegraphics[scale=.4]{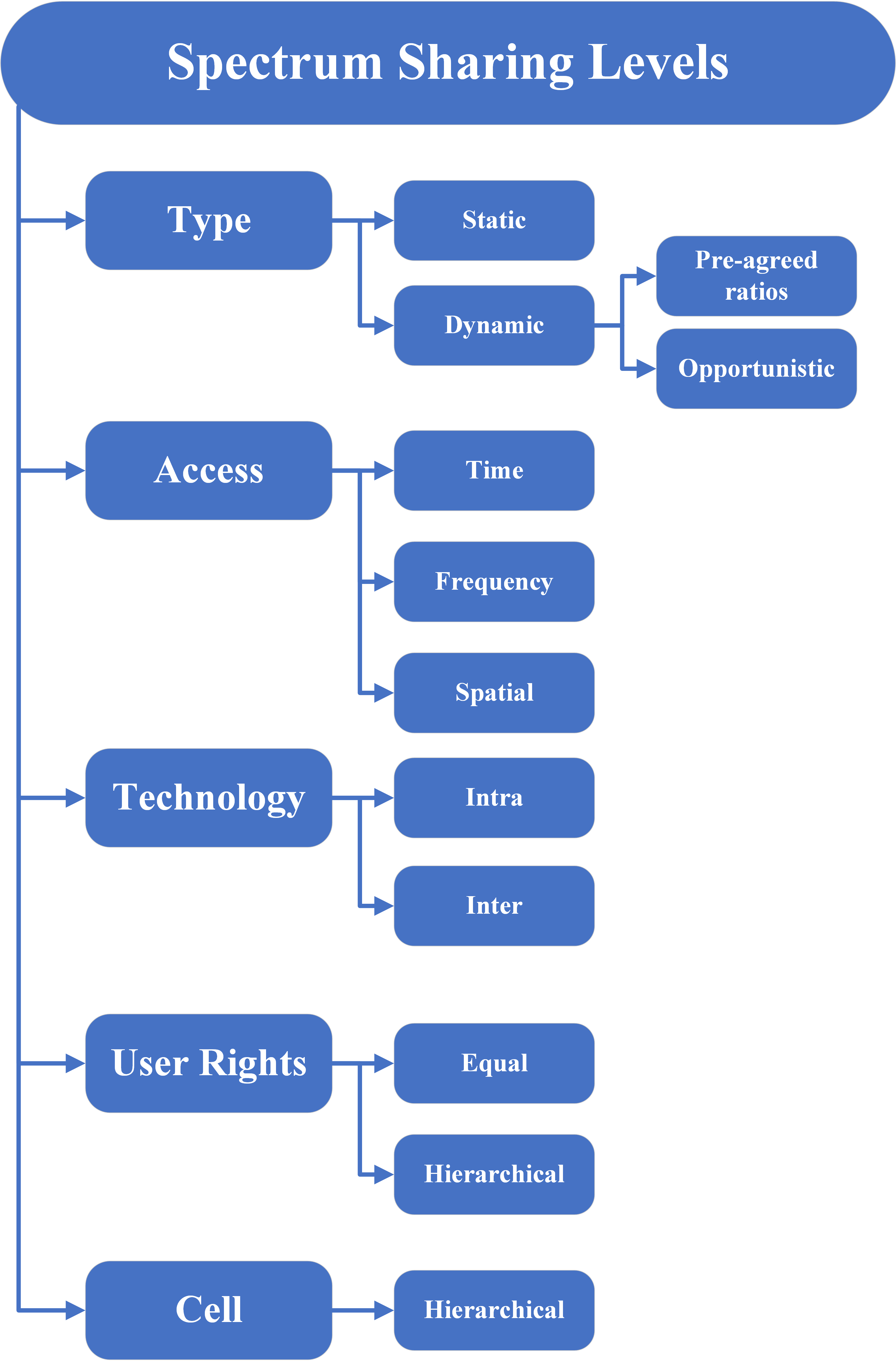}
	\caption{Different Levels of Spectrum Sharing}
	\label{spectrum_sharing_decomposition}
\end{figure}

\indent As mentioned earlier, spectrum sharing has been proposed as one potential solution to the spectrum under-utilization problem. It allows users/technologies/applications to re-use or share the radio resources to improve the spectrum efficiency and increase the throughput without needing new resources. The spectrum sharing process can be implemented at different levels and using different techniques as illustrated in Fig. \ref{spectrum_sharing_decomposition}. In what follows, some of the different levels and mechanisms for spectrum sharing are presented along with a few related previous works. 
\subsection{Spectrum Sharing Types}
\indent The first level of spectrum sharing to consider is its type. In essence, this describes the manner in which the spectrum is shared. Accordingly, spectrum can be shared in a static or dynamic manner. When static spectrum sharing is applied, a pre-defined portion of the shared spectrum is assigned to a particular user or SP. As an example, assume 2 SPs each have 5 resource blocks (RBs) within an LTE system. These RBs can be grouped to one larger pool consisting of 10 RBs. Under a static spectrum sharing model, the first SP is always allocated the first 7 RBs to assign to its users while the second SP is allocated the last 3 RBs as per a service level agreement (SLA) between them. Although this might mimic the traditional licensing-based allocation model previously adopted, it still constitutes a spectrum sharing component given that one SP is using some of the RBs licensed to a second SP. Using this model, Hussein \textit{et al.} proposed a static spectrum sharing mechanism for an uplink LTE system \cite{Moubayed_WRV_pa} by formulating a binary integer programming problem with the objective being to minimize the overall transmission power for all users. The authors proposed the mechanisms and investigated the resulting power savings if SPs shared their resources based on a pre-agreed SLA ratio with results showing that the proposed mechanism does indeed reduce the transmission power by allowing access to a larger pool of RBs for the users.\\
\indent In contrast, dynamic spectrum sharing allows more flexibility in terms of the allocation by allowing users or SPs to access any of the resources within the shared spectrum. This can be done either based on a pre-agreed sharing ratio or it can be done opportunistically. In the pre-agreed sharing ratio scenario, users or SPs are allocated a portion of the resouces available based on the SLA agreement. However, the resources allocated may differ between two time instances. Using the same example as above, when adopting a dynamic spectrum sharing model under a pre-agreed sharing ratio scenario, the first SP would again be allocated 7 RBs while the second SP would get 3 RBs. However, the main difference in this case is the fact that the order of RBs allocated to each SP is irrelevant and can change between one time instance and another. For example, at time instance $t=t_1$, the first SP can be allocated RBs [1,2,4,5,7,9,10] while the second SP would get RBs [3,6,8]. This can change at time instance $t=t_2$ in which the first SP is allocated RBs [2,3,5,6,8,9,10] while the second SP is allocated RBs [1,4,7]. Accordinglt, such a model adds more flexibility and provides better opportunities to the users or SPs given that they are not bound by a rigid set of resources. Kalil \textit{et al.} adopted a dynamic spectrum sharing model with pre-agreed sharing ratios \cite{Moubayed_WRV_fairness,Kalil_TMC}. In their work, the authors proposed a dynamic spectrum sharing optimization model coupled with an efficient low-complexity scheduler to share the LTE RBs between users of multiple SPs. The proposed scheduler maximized the throughput while maintaining access proportional fairness among users as well as SPs. Simulation results showed that the scheduler was able to improve the average aggregate throughput while ensuring the fairness between users and SPs.\\
\indent On the other hand, opportunistic dynamic spectrum sharing refers to the concept of unlicensed users using idle radio resources owned by a licensed SP \cite{opportunistic_spectrum_sharing1}. This can be thought of as adopting a ``Spectrum-as-a-Service'' architecture in which users or SPs can get access to radio resources ``on-demand''. This type of dynamic spectrum sharing is prevalent in cognitive radio networks which rely on unlicensed users (commonly referred to as secondary users) to sense and detect spectrum gaps and ``opportunistically'' use these gaps \cite{Moubayed_CR1,Moubayed_CR2}. Li \textit{et al.} proposed such a spectrum sharing model \cite{opportunistic_spectrum_sharing1}. The authors derived  a tight upper bound on the bit error rate (BER) and showed that the proposed opportunistic spectrum access-based model achieved lower BER and a higher spectral efficiency \cite{opportunistic_spectrum_sharing1}. In a similar fashion, Jiang and Mao proposed an inter-operator opportunistic spectrum sharing model in LTE-unlicensed for LTE base stations that guarantees users' quality of service requirements \cite{opportunistic_spectrum_sharing2}. Simulation results showed that the proposed model achieved the lowest average packet drop rate, lowest average packet delay, and the highest average throughput \cite{opportunistic_spectrum_sharing2}. This further illustrates the merits of the opportunistic dynamic spectrum sharing/Spectrum-as-a-Service model/architecture.   
\subsection{Spectrum Sharing Access}
\indent A second level at which the spectrum sharing process can be discussed at is the access mechanisms. This refers to how the different users/SPs/technologies share the provided radio resources \cite{spectrum_sharing3}. Multiple spectrum sharing access mechanisms have been proposed in the literature including using time, frequency, spatial, or code division multiple access \cite{spectrum_sharing3}. These mechanisms allow spectrum sharing using different domains.\\
\indent One intuitive spectrum sharing access mechanism is in the time domain. This is commonly referred to as time division multiple access (TDMA) \cite{TDMA1}. Simply put, when adopting such a mechanism, each user/SP/technology uses a different time slot to communicate \cite{TDMA1}. More specifically, the user/SP/technology has access to the entire spectrum resources during the allocated time slot. The allocation of this time slot can be either done in a static (\textit{i.e.} a user/SP/technology is allocated the same time slot in each scheduling window in a round robin manner) or dynamic (\textit{i.e} a user/SP/technology is allocated different time slots in each scheduling window) manner \cite{TDMA1}. The static TDMA mechanism is easier to implement and manage, but is considered to be less spectrum efficient. In contrast, adopting a dynamic TDMA mechanism adds more flexibility and results in better spectrum efficiency at the expense of additional implementation and management complexity. One example of a TDMA-based spectrum sharing mechanism was proposed by Hu \textit{et al.} \cite{TDMA1}. In their work, the authors assumed the opportunistic access of secondary users in a cognitive radio network to unused time slots \cite{TDMA1}. Their numerical results showed that the proposed scheme improved the spectrum utilization without impact the primary users' transmission and performance \cite{TDMA1}.\\
\indent A second spectrum sharing access mechanism is in the frequency domain and it is commonly referred to as frequency division multiple access (FDMA) \cite{FDMA1}. Similar to its time counterpart, this sharing mechanism assumes that a user/SP/technology uses a different frequency band at any moment in time \cite{FDMA1,dechene}. Again this can be done in a static (\textit{i.e.} a user/SP/technology is always assigned the same frequency band or resource in each scheduling window) or dynamic (\textit{i.e.} a user/SP/technology is assigned different frequency bands or resources in each scheduling window) manner. Kalil \textit{et al.} proposed a dynamic frequency-based spectrum sharing access mechanism \cite{Kalil_WRV}. Their work assumed that the original spectrum assigned to three LTE SPs was combined into one large spectrum consisting of the aggregate RBs of the three SPs. The RBs were then allocated to the users while respecting the SLA ratio agreed on previously between the SPs. Simulation results showed that the proposed mechanisms improved the spectrum utilization and efficiency highlighted by the higher resulting throughput (due to users having access to a larger pool of RBs) while maintaining the SLA ratios defined \cite{Kalil_WRV,Kalil_TVT}.\\
\indent A third spectrum sharing mechanism is in the spatial domain (SDMA). When using this mechanisms, users or base stations use beam-forming techniques to direct the communication and limit it to a defined direction rather than being omni-directional \cite{SDMA1}. Despite the complexity of incorporating the beam-forming techniques, the benefit of such a mechanism is that it reduces mutual interference and allows different users to access the entire spectrum simultaneously which in turn results in a better spectrum utilization \cite{SDMA1}. Accordingly, Feng \textit{et al.} proposed a distributed SDMA-based spectrum sharing mechanism to eliminate mutual interference between coexisting systems in overlapping areas \cite{SDMA1}. The authors' simulation results showed that the proposed mechanisms significantly reduced the mutual interference and improved the spectrum efficiency.
\subsection{Spectrum Sharing at Technology Level}
\indent A third level at which spectrum sharing can be discussed is at the technology level. As the name suggests, this refers to the process in which users belonging to either the same technology or different technologies share the spectrum. The former is commonly referred to as intra-technology while the latter is referred to as inter-technology spectrum sharing. The benefit of discussing spectrum sharing at the technology level is that it can include various spectrum sharing mechanisms both in terms of access and sharing type as discussed earlier.\\
\indent Several previous works have proposed spectrum sharing at the intra-technology level. More specifically, Moubayed \textit{et al.} proposed an LTE/LTE-A system in which both cellular and device-to-device (D2D) users belonging to multiple SPs share the licensed LTE spectrum \cite{Moubayed_WRV_D2D,Moubayed_WRV_D2D_pa}. In their work, the authors assumed that the RBs of these SPs are aggregated into one larger pool available for the users to dynamically share while still maintaining SLA and throughput requirements. The authors also investigated the transmission power reduction achievable due to the dynamic sharing of the resources. The author further extended their work to the case of machine-to-machine communication and showed that the resulting dynamic spectrum sharing process achieves higher average throughputs while still guaranteeing QoS and SLA requirements \cite{Moubayed_WRV_M2M}.\\
\indent In contrast, Zimmo \textit{et al.} proposed an inter-technology spectrum sharing between LTE and WiFi in the unlicensed band using both time and frequency domain \cite{Moubayed_WRV_WiFi}. More specifically, the authors proposed a static time-based spectrum sharing between the LTE and WiFi technologies in which a dedicated time slot is given to each technology using different configurations \cite{Moubayed_WRV_WiFi}. On the other hand, the work assumed a dynamic spectrum sharing process between the users of each technology with LTE users having access to a larger pool of RBs to be allocated according to SLA requirements and WiFi users being able to connect to access points belonging to other SPs \cite{Moubayed_WRV_WiFi}. Simulation results showed that the average throughput of each technology improved due to the users having access to better channels due to the dynamic spectrum sharing process \cite{Moubayed_WRV_WiFi}.
\subsection{Spectrum Sharing User Rights}
\indent A fourth level that can be used to characterize the spectrum sharing process is the user access rights level. This pertains to whether different users have equal access rights when sharing the spectrum or whether one is given priority over the other. This is important given the potential impact this has on the overall system performance, particularly in terms of the possible interference caused when sharing the spectrum.\\
\indent Several research works have proposed an equal user access rights spectrum sharing mechanism. One such example is the work by Poulakis \textit{et al.} \cite{equal_user_spectrum_sharing}. The authors investigated the potential of allowing 5G and D2D technologies to co-exist with the D2D being used to partially offload the 5G cellular traffic. Simulation results showed that such a mechanism would increase the overall network throughput on the condition that the cellular base stations' density is not increased. These results are in line with the results in \cite{Moubayed_WRV_D2D} and \cite{Moubayed_WRV_M2M} in which it was also shown that introducing D2D and M2M communication while controlling the interference they caused did indeed increase the overall throughput of the network. \\
\indent On the other hand, multiple researchers proposed the use of a hierarchical user access rights framework for spectrum sharing in which priority is given to one group of users over the other. This is common when adopting a cognitive radio network framework in which priority is typically given to primary users (PUs) over secondary users (SUs). One work that adopted a hierarchical user access rights framework is Saki \textit{et al.} \cite{hierarchical_user_spectrum_sharing1}. The authors assumed that SUs can opportunistically detect spectrum gaps and utilize them (in overlay mode) as well as transmit at lower power levels when PUs are active (in underlay mode) \cite{hierarchical_user_spectrum_sharing1}. The work proposed the use of stochastic transmit and interference power constraints to ensure that the performance of PUs is protected \cite{hierarchical_user_spectrum_sharing1}. Similarly, Thakur \textit{et al.} also proposed a combined overlay-underlay spectrum access framework in which SUs transmit at regular power when PUs are idle and at reduced power when PUs are active \cite{hierarchical_user_spectrum_sharing2}. The authors derived the closed-form expressions for both the throughput and data loss in such a system. Using simulations, the authors showed that the proposed framework improved the throughput and reduced data loss for PUs when SUs are active when compared to the conventional approaches. This highlights the benefits of adopting a dynamic spectrum sharing process through equal/hierarchical user access rights frameworks.
\subsection{Spectrum Sharing at Cell Level}
\indent Another level at which spectrum can be shared is at the cell level. This is common when a multi-tier cellular heterogeneous network architecture consisting of cells of different sizes is adopted \cite{multi_cell_spectrum_sharing1,multi_cell_spectrum_sharing2}. Within such an architecture, cells of different sizes ranging from macro-cells to micro-cells all the way to pico-cells and femto-cells can share the same spectrum. The spectrum sharing in this case is often governed by the transmission power level with larger cells using larger levels \cite{multi_cell_spectrum_sharing1,multi_cell_spectrum_sharing2}. Hussein \textit{et al.} proposed spectrum sharing at the cell level in \cite{Moubayed_WRV_5G}. The authors proposed a dynamic spectrum sharing scheme in which macro-cell users are able to share the LTE resources available at the micro-cell whenever they are within the micro-cell's range \cite{Moubayed_WRV_5G}. Simulation results showed that such a multi-tier spectrum sharing scheme allows macro-cell users to achieve lower average delay due to the temporary association with the shorter range micro-cell base station \cite{Moubayed_WRV_5G}. This illustrates the significant benefits of adopting a dynamic spectrum sharing scheme.\\
\section{Machine Learning}\label{ML}
\indent Adopting dynamic spectrum sharing and enabling a ``Spectrum-as-a-Service'' architecture results in a multitude of benefits such as higher network throughputs, lower average latencies, improved spectrum utilization, and reduced capital and operational expenditure. However, adopting such architectures and mechanisms comes with a new set of challenges such as when to make the sharing decision, in what capacity, and to what extent. Many of the previous works in the literature formulated mathematical optimization models (such as integer linear programming, mixed integer linear programming, and mixed integer nonlinear programming models)  to address these questions. However, such models are associated with high computational complexity as shown in \cite{Moubayed_WRV_D2D,Moubayed_WRV_D2D_pa,Moubayed_WRV_M2M}. This makes them unsuitable for real-time decision making and adoption. One potential solution is ML techniques and algorithms. Such algorithms can extract useful information from data collected about the behavior and characteristics of the system without explicit programming \cite{ML1,ML2}. Different models belonging to different ML algorithm categories including supervised learning, unsupervised learning, semi-supervised learning, deep learning, and reinforcement learning (RL) \cite{Moubayed_ML} can be adapted to help automate and facilitate the process of dynamic spectrum sharing and offer an improved Spectrum-as-a-Service architecture. 
\subsection{Brief Overview}
\indent Starting with the first category, supervised learning is the group of algorithms in which the learning process is completed using a labeled training dataset \cite{Moubayed_ML}. The goal is to predict an output value $y_{new}$ for a particular input vector $x_{new}$ based on a function learned using a group of training pairs $(x,y)$. In contrast, unsupervised learning is the group of algorithms in which the learning process is completed using an unlabeled training dataset \cite{Moubayed_ML}. In this case, the goal is to determine a pattern by grouping similar points based on some similarity metric (typically a distance metric such as euclidean distance or Manhattan distance) \cite{Moubayed_ML}. The third category, namely semi-supervised learning, combines aspects from the previous two categories. Therefore, it aims at learning the function or pattern using a partially labeled training dataset \cite{Moubayed_ML}. Deep learning algorithms are essentially large-scale neural networks whose goal is to model data abstractions using a graph of multiple processing layers made up of units called neurons \cite{Moubayed_ML}. These neurons apply linear and non-linear transformations using what is known as activation functions on the input data to extract useful information from it \cite{Moubayed_ML}. Finally, RL algorithms follow a trial-and-error methodology by taking an action with the goal of maximizing a cumulative reward metric \cite{Moubayed_ML}. These diverse set of algorithms provide a vast array of opportunities for adaptation to automate and enable efficient, effective, and robust dynamic spectrum sharing mechanisms and Spectrum-as-a-Service architectures.
\subsection{Previous Efforts}
\indent The beauty of adopting ML algorithms as key enablers for dynamic spectrum sharing is that they can be applied at different levels or stages of the spectrum sharing process starting with the channel sensing and condition prediction stage all the way to the spectrum sharing decision stage. As such, multiple works from the literature proposed the use of ML techniques to enable dynamic spectrum sharing \cite{ML_spectrum_sharing1,ML_spectrum_sharing2,ML_spectrum_sharing3}.\\
\indent Rastegardoost \textit{et al.} proposed an RL-based model to enable the dynamic spectrum sharing between WiFi and LTE in the unlicensed band \cite{ML_spectrum_sharing1}. More specifically, the authors used the Q-learning model to offer an online distributive robust and model-free decision-making framework for WiFi and LTE coexistence \cite{ML_spectrum_sharing1}. The goal is to minimize the latency experience by WiFi users while maximizing the idle resource utilization by the LTE users \cite{ML_spectrum_sharing1}. Simulation results showed that the proposed RL-based model significantly reduced the WiFi latency when compared to the Almost Blank Subframe (ABS) approach while achieving comparable performance in terms of LTE throughput. This illustrates the potential of using ML algorithms to enable the efficient coexistence between technologies and facilitate dynamic spectrum sharing among them.\\
\indent Similarly, Jiang \textit{et al.} proposed the use of RL techniques in a cognitive radio network environment \cite{ML_spectrum_sharing2}. The authors introduced two novel RL-based approaches for the efficient exploration and exploitation of the spectrum. Moreover, the authors investigated and derived the learning efficiency of these approaches within the considered environment \cite{ML_spectrum_sharing2}. Simulation results showed that the proposed approaches significantly improved the learning efficiency by reducing the number of trials needed by the learning agent to learn each task. Furthermore, it was shown that using these approaches reduced the blocking and dropping probability compared to traditional spectrum sharing mechanisms in cognitive radio environments \cite{ML_spectrum_sharing2}.
\subsection{Research Opportunities}
\begin{figure}[!h]
	\centering
	\includegraphics[scale=.45]{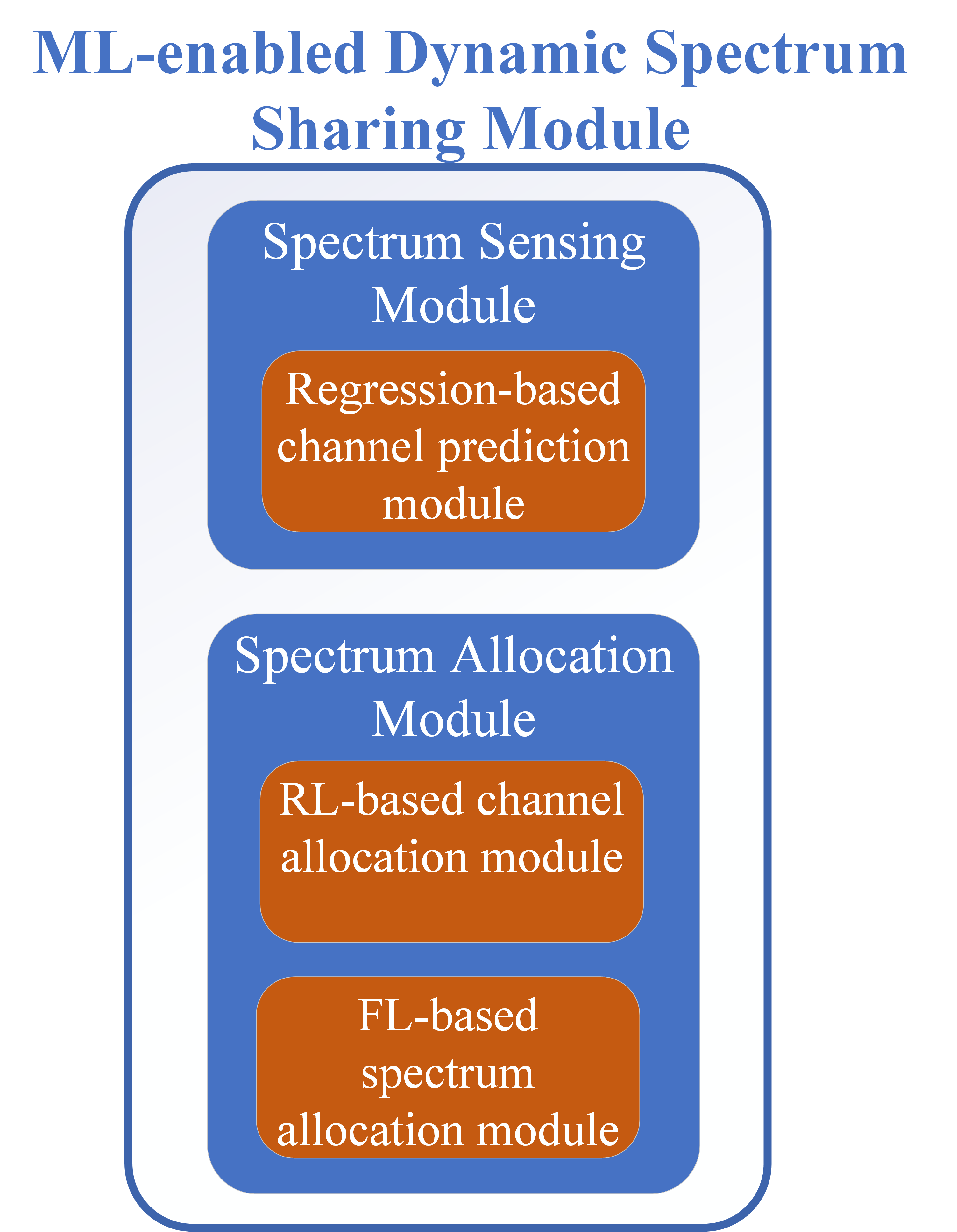}
	\caption{Potential ML-based Dynamic Spectrum Sharing Module}
	\label{spectrum_sharing_ML_fig}
\end{figure}

\indent As shown above, ML techniques and algorithms have illustrated their potential in improving the spectrum sharing process in a bid to offer more flexible and dynamic wireless communication systems. However, there are more opportunities in which ML can play an extensive role in further enhancing dynamic spectrum sharing. \\
\indent One research opportunity is applying different regression models such as polynomial regression of support vector regression (SVR) for spectrum sensing across multiple frequency bands. Similar in concept to the work in \cite{ML_spectrum_sharing3}, the goal is to predict the channel conditions for the different users across multiple technologies and spectrum bands. Having an accurate prediction of the channel conditions can then be used as a prelude to making channel selection/allocation and transmission power decisions. Therefore, different SPs can develop various channel condition prediction regression models deployed at the base stations and communication using the data collected from their users such as spectrum occupancy data, device nonlinearity information, transmitted and received power information, and detection of abnormal signals such as interference. Using this information, new user requests can be allocated to good quality channels in different frequency bands, providing a flexible on-demand Spectrum-as-a-Service architecture.\\
\indent Another opportunity is extending the concept of RL proposed in \cite{ML_spectrum_sharing2} to all frequency spectrum bands rather than just for a cognitive radio environment. In this case, the RL agent will learn over time what decisions result in a higher reward. Moreover, this agent can use the predicted channel conditions from the regression models discussed above as part of the system input to it when making channel allocation decisions. Techniques such as Q-learning or Proximal Policy Optimization can be used as part of the RL framework that would receive user spectrum requests and allocate suitable spectrum channels.\\
\indent A third opportunity is adopted the concept of federated learning (FL). FL is a novel ML paradigm that provides a centralized model trained using data distributed across multiple locations \cite{FL1,FL2}. This paradigm essentially brings the code to the data rather than bringing the data to the code, thereby addressing potential concerns regarding the data privacy, ownership, and locality \cite{FL2}. Using such a paradigm, data collected at different base stations (such as spectrum occupancy data, user congestion data, transmitted and received power information, and detected interference signals) can then be used to train a local spectrum sharing model whose parameters are then sent back to the core network. The parameters are then aggregated to find a global spectrum sharing ML model that is shared back to the base stations to implement. The advantage of adopting the FL paradigm is that it allows different technologies to participate in the model learning process without jeopardizing their users' privacy. Fig. \ref{spectrum_sharing_ML_fig} summarizes these opportunities.
\section{Conclusion}\label{conc}
\indent A significant demand increase for new radio frequency spectrum has been observed due to the growth in wireless broadband users, devices, and applications. This is expected to grow further given the projection that the global traffic per year will reach 4.8 zettabytes by 2022 and the number of connected devices will be close 28.5 billion devices \cite{Cisco_data_growth}. However, providing more spectrum to expand existing services or offer new ones has become more challenging due to the spectrum being mostly allocated \cite{us_freq_allocation_2019}. To address this, spectrum sharing has been proposed as a potential solution to improve spectrum utilization. Given that spectrum sharing can be adopted at different levels, adopting effective and efficient spectrum sharing mechanisms is in itself challenging. To that end, this paper provided a brief overview of the different spectrum sharing levels and techniques previously proposed in the literature. Moreover, it illustrated the benefits of adopting dynamic sharing mechanisms by offering Spectrum-as-a-Service architecture that allows for higher spectrum utilization rates, higher network throughputs, and lower packet delays. Furthermore, it described the potential role of ML models and paradigms such as regression, RL, and FL in facilitating the efficient dynamic sharing of the spectrum and offering Spectrum-as-a-Service.

\small
\bibliographystyle{IEEEtran}
\bibliography{Ref1}

\begin{thebibliography}{10}
\providecommand{\url}[1]{#1}
\csname url@samestyle\endcsname
\providecommand{\newblock}{\relax}
\providecommand{\bibinfo}[2]{#2}
\providecommand{\BIBentrySTDinterwordspacing}{\spaceskip=0pt\relax}
\providecommand{\BIBentryALTinterwordstretchfactor}{4}
\providecommand{\BIBentryALTinterwordspacing}{\spaceskip=\fontdimen2\font plus
\BIBentryALTinterwordstretchfactor\fontdimen3\font minus
  \fontdimen4\font\relax}
\providecommand{\BIBforeignlanguage}[2]{{%
\expandafter\ifx\csname l@#1\endcsname\relax
\typeout{** WARNING: IEEEtran.bst: No hyphenation pattern has been}%
\typeout{** loaded for the language `#1'. Using the pattern for}%
\typeout{** the default language instead.}%
\else
\language=\csname l@#1\endcsname
\fi
#2}}
\providecommand{\BIBdecl}{\relax}
\BIBdecl

\bibitem{Cisco_data_growth}
{Cisco}, ``{Cisco Predicts More IP Traffic in the Next Five Years Than in the
  History of the Internet},'' Nov. 2018.

\bibitem{us_freq_allocation_2019}
{Federal Communications Commission}, ``Fcc online table of frequency
  allocations,'' Tech. Rep., 2020.

\bibitem{spectrum_inefficiency1}
P.~Tilghman, ``Will rule the airwaves: A darpa grand challenge seeks autonomous
  radios to manage the wireless spectrum,'' \emph{IEEE Spectrum}, vol.~56,
  no.~6, pp. 28--33, 2019.

\bibitem{spectrum_inefficiency2}
G.~L. Rosston, ``Increasing the efficiency of spectrum allocation,''
  \emph{Springer's Review of Industrial Organization}, vol.~45, no.~3, pp.
  221--243, 2014.

\bibitem{5G_requirements}
T.~Norp, ``5g service requirements,'' \emph{3gpp. org}, 2017.

\bibitem{spectrum_sharing1}
L.~{Zhang}, M.~{Xiao}, G.~{Wu}, M.~{Alam}, Y.~{Liang}, and S.~{Li}, ``A survey
  of advanced techniques for spectrum sharing in 5g networks,'' \emph{IEEE
  Wireless Communications}, vol.~24, no.~5, pp. 44--51, 2017.

\bibitem{spectrum_sharing2}
F.~{Hu}, B.~{Chen}, and K.~{Zhu}, ``Full spectrum sharing in cognitive radio
  networks toward 5g: A survey,'' \emph{IEEE Access}, vol.~6, pp.
  15\,754--15\,776, 2018.

\bibitem{spectrum_sharing3}
A.~M. {Voicu}, L.~{Simić}, and M.~{Petrova}, ``Survey of spectrum sharing for
  inter-technology coexistence,'' \emph{IEEE Communications Surveys \&
  Tutorials}, vol.~21, no.~2, pp. 1112--1144, 2019.

\bibitem{spectrum_sharing4}
R.~H. {Tehrani}, S.~{Vahid}, D.~{Triantafyllopoulou}, H.~{Lee}, and
  K.~{Moessner}, ``Licensed spectrum sharing schemes for mobile operators: A
  survey and outlook,'' \emph{IEEE Communications Surveys \& Tutorials},
  vol.~18, no.~4, pp. 2591--2623, 2016.

\bibitem{spectrum_sharing_cost_benefit}
H.~Zhou, Q.~Yu, X.~S. Shen, S.~Wu, and Q.~Zhang, ``Dynamic wireless spectrum
  sharing in cognitive cellular networks,'' in \emph{Dynamic Sharing of
  Wireless Spectrum}.\hskip 1em plus 0.5em minus 0.4em\relax New York:
  Springer, 2017, pp. 37--57.

\bibitem{spectrum_sharing_challenges}
S.~{Bhattarai}, J.~J. {Park}, B.~{Gao}, K.~{Bian}, and W.~{Lehr}, ``An overview
  of dynamic spectrum sharing: Ongoing initiatives, challenges, and a roadmap
  for future research,'' \emph{IEEE Transactions on Cognitive Communications
  and Networking}, vol.~2, no.~2, pp. 110--128, 2016.

\bibitem{ML_networking}
M.~{Wang}, Y.~{Cui}, X.~{Wang}, S.~{Xiao}, and J.~{Jiang}, ``{Machine Learning
  for Networking: Workflow, Advances and Opportunities},'' \emph{{IEEE
  Network}}, vol.~32, no.~2, pp. 92--99, Mar. 2018.

\bibitem{ML_spectrum_sharing1}
N.~Rastegardoost and B.~Jabbari, ``A machine learning algorithm for unlicensed
  lte and wifi spectrum sharing,'' in \emph{2018 IEEE International Symposium
  on Dynamic Spectrum Access Networks (DySPAN)}.\hskip 1em plus 0.5em minus
  0.4em\relax IEEE, Seoul, South Korea, Oct. 2018, pp. 1--6.

\bibitem{ML_spectrum_sharing2}
T.~Jiang, D.~Grace, and P.~D. Mitchell, ``Efficient exploration in
  reinforcement learning-based cognitive radio spectrum sharing,'' \emph{IET
  communications}, vol.~5, no.~10, pp. 1309--1317, 2011.

\bibitem{ML_spectrum_sharing3}
Z.~{Zhang}, K.~{Zhang}, F.~{Gao}, and S.~{Zhang}, ``Spectrum prediction and
  channel selection for sensing-based spectrum sharing scheme using online
  learning techniques,'' in \emph{2015 IEEE 26th Annual International Symposium
  on Personal, Indoor, and Mobile Radio Communications (PIMRC)}, Hong Kong,
  China, Aug. 2015, pp. 355--359.

\bibitem{ML_statistic1}
L.~Columbus, ``Roundup of machine learning forecast and market estimates,
  2018,'' \emph{{Forbes Magazine}}, 2018.

\bibitem{Moubayed_WRV_pa}
M.~{Hussein}, A.~{Moubayed}, S.~{Primak}, and A.~{Shami}, ``On efficient power
  allocation modeling in virtualized uplink 3gpp-lte systems,'' in \emph{2015
  IEEE 11th International Conference on Wireless and Mobile Computing,
  Networking and Communications (WiMob)}, Abu Dhabi, UAE, Oct. 2015, pp.
  817--824.

\bibitem{Moubayed_WRV_fairness}
M.~{Kalil}, A.~{Moubayed}, A.~{Shami}, and A.~{Al-Dweik}, ``Efficient
  low-complexity scheduler for wireless resource virtualization,'' \emph{IEEE
  Wireless Communications Letters}, vol.~5, no.~1, pp. 56--59, 2016.

\bibitem{Kalil_TMC}
M.~{Kalil}, A.~{Shami}, and A.~{Al-Dweik}, ``Qos-aware power-efficient
  scheduler for lte uplink,'' \emph{IEEE Transactions on Mobile Computing},
  vol.~14, no.~8, pp. 1672--1685, 2015.

\bibitem{opportunistic_spectrum_sharing1}
Q.~{Li}, M.~{Wen}, S.~{Dang}, E.~{Basar}, H.~V. {Poor}, and F.~{Chen},
  ``Opportunistic spectrum sharing based on ofdm with index modulation,''
  \emph{IEEE Transactions on Wireless Communications}, vol.~19, no.~1, pp.
  192--204, 2020.

\bibitem{Moubayed_CR1}
A.~{Moubayed}, S.~{Sorour}, T.~{Al-Naffouri}, and M.~{Alouini}, ``Collaborative
  multi-layer network coding in hybrid cellular cognitive radio networks,'' in
  \emph{2015 IEEE 81st Vehicular Technology Conference (VTC Spring)}, Glasgow,
  Scotland, May 2015, pp. 1--6.

\bibitem{Moubayed_CR2}
\BIBentryALTinterwordspacing
A.~Moubayed, ``Collaborative multi-layer network coding for hybrid cellular
  cognitive radio networks,'' Master's thesis, 2014. [Online]. Available:
  \url{http://hdl.handle.net/10754/316697}
\BIBentrySTDinterwordspacing

\bibitem{opportunistic_spectrum_sharing2}
Z.~{Jiang} and S.~{Mao}, ``Interoperator opportunistic spectrum sharing in
  lte-unlicensed,'' \emph{IEEE Transactions on Vehicular Technology}, vol.~66,
  no.~6, pp. 5217--5228, 2017.

\bibitem{TDMA1}
S.~Hu, Y.-D. Yao, and Z.~Yang, ``Cognitive medium access control protocols for
  secondary users sharing a common channel with time division multiple access
  primary users,'' \emph{Wiley's Wireless Communications and Mobile Computing},
  vol.~14, no.~2, pp. 284--296, 2014.

\bibitem{FDMA1}
B.~Da and C.~Ko, ``Dynamic spectrum sharing in orthogonal frequency division
  multiple access--based cognitive radio,'' \emph{IET communications}, vol.~4,
  no.~17, pp. 2125--2132, 2010.

\bibitem{dechene}
D.~J. {Dechene} and A.~{Shami}, ``Energy-aware resource allocation strategies
  for lte uplink with synchronous harq constraints,'' \emph{IEEE Transactions
  on Mobile Computing}, vol.~13, no.~2, pp. 422--433, 2014.

\bibitem{Kalil_WRV}
M.~Kalil, A.~Shami, and Y.~Ye, ``Wireless resources virtualization in lte
  systems,'' in \emph{2014 IEEE Conference on Computer Communications Workshops
  (INFOCOM WKSHPS)}.\hskip 1em plus 0.5em minus 0.4em\relax IEEE, Toronto,
  Canada, Apr. 2014, pp. 363--368.

\bibitem{Kalil_TVT}
M.~{Kalil}, A.~{Shami}, A.~{Al-Dweik}, and S.~{Muhaidat}, ``Low-complexity
  power-efficient schedulers for lte uplink with delay-sensitive traffic,''
  \emph{IEEE Transactions on Vehicular Technology}, vol.~64, no.~10, pp.
  4551--4564, 2015.

\bibitem{SDMA1}
Y.~Feng, S.~Yang, Q.~Jigang, and X.~Binyang, ``Distributed spatial division
  multiple access technique for spectrum sharing systems,'' \emph{Bell Labs
  Technical Journal}, vol.~13, no.~4, pp. 119--128, 2009.

\bibitem{Moubayed_WRV_D2D}
A.~{Moubayed}, A.~{Shami}, and H.~{Lutfiyya}, ``Wireless resource
  virtualization with device-to-device communication underlaying lte network,''
  \emph{IEEE Transactions on Broadcasting}, vol.~61, no.~4, pp. 734--740, 2015.

\bibitem{Moubayed_WRV_D2D_pa}
------, ``Power-aware wireless virtualized resource allocation with d2d
  communication underlaying lte network,'' in \emph{2016 IEEE Global
  Communications Conference (GLOBECOM)}, Washington DC, USA, Dec. 2016, pp.
  1--6.

\bibitem{Moubayed_WRV_M2M}
A.~{Moubayed}, K.~{Hammad}, A.~{Sham}, and H.~{Lutfiyya}, ``Dynamic spectrum
  management through resource virtualization with m2m communications,''
  \emph{IEEE Communications Magazine}, vol.~56, no.~10, pp. 121--127, 2018.

\bibitem{Moubayed_WRV_WiFi}
S.~{Zimmo}, A.~{Moubayed}, A.~{Refaey}, and A.~{Shami}, ``Coexistence of wifi
  and lte in the unlicensed band using time-domain virtualization,'' in
  \emph{2018 IEEE Global Communications Conference (GLOBECOM)}, Abu Dhabi, UAE,
  Dec. 2018, pp. 1--6.

\bibitem{equal_user_spectrum_sharing}
M.~I. Poulakis, A.~G. Gotsis, and A.~Alexiou, ``Multicell device-to-device
  communication: A spectrum-sharing and densification study,'' \emph{IEEE
  Vehicular Technology Magazine}, vol.~13, no.~1, pp. 85--96, 2018.

\bibitem{hierarchical_user_spectrum_sharing1}
H.~Saki, A.~Shojaeifard, and M.~G. Martini, ``Stochastic resource allocation
  for hybrid spectrum access ofdma-based cognitive radios,'' in \emph{2015 IEEE
  International Conference on Communications (ICC)}.\hskip 1em plus 0.5em minus
  0.4em\relax IEEE, London, UK, Jun. 2015, pp. 7750--7755.

\bibitem{hierarchical_user_spectrum_sharing2}
P.~Thakur, A.~Kumar, S.~Pandit, G.~Singh, and S.~Satashia, ``Advanced frame
  structures for hybrid spectrum access strategy in cognitive radio
  communication systems,'' \emph{IEEE Communications Letters}, vol.~21, no.~2,
  pp. 410--413, 2016.

\bibitem{multi_cell_spectrum_sharing1}
C.~{Yang}, J.~{Li}, M.~{Guizani}, A.~{Anpalagan}, and M.~{Elkashlan},
  ``Advanced spectrum sharing in 5g cognitive heterogeneous networks,''
  \emph{IEEE Wireless Communications}, vol.~23, no.~2, pp. 94--101, 2016.

\bibitem{multi_cell_spectrum_sharing2}
E.~{Hossain}, M.~{Rasti}, H.~{Tabassum}, and A.~{Abdelnasser}, ``Evolution
  toward 5g multi-tier cellular wireless networks: An interference management
  perspective,'' \emph{IEEE Wireless Communications}, vol.~21, no.~3, pp.
  118--127, 2014.

\bibitem{Moubayed_WRV_5G}
M.~{Hussein}, A.~{Moubayed}, S.~{Primak}, and A.~{Shami}, ``Virtualized
  allocation performance analysis in 5g two-tier cellular networks,'' in
  \emph{2016 IEEE Canadian Conference on Electrical and Computer Engineering
  (CCECE)}, Vancouver, Canada, May 2016, pp. 1--4.

\bibitem{ML1}
R.~Schapire, ``{COS 511: Theoretical Machine Learning},'' 2008, {Princeton
  University}.

\bibitem{ML2}
A.~Rostamizadeh and A.~Talwalkar, \emph{Foundations of {Machine
  Learning}}.\hskip 1em plus 0.5em minus 0.4em\relax MIT Press.

\bibitem{Moubayed_ML}
A.~{Moubayed}, M.~{Injadat}, A.~B. {Nassif}, H.~{Lutfiyya}, and A.~{Shami},
  ``E-learning: Challenges and research opportunities using machine learning \&
  data analytics,'' \emph{IEEE Access}, vol.~6, pp. 39\,117--39\,138, 2018.

\bibitem{FL1}
J.~Kone{\v{c}}n{\`y}, H.~B. McMahan, F.~X. Yu, P.~Richt{\'a}rik, A.~T. Suresh,
  and D.~Bacon, ``Federated learning: Strategies for improving communication
  efficiency,'' \emph{arXiv preprint arXiv:1610.05492}, 2016.

\bibitem{FL2}
K.~Bonawitz, H.~Eichner, W.~Grieskamp, D.~Huba, A.~Ingerman, V.~Ivanov,
  C.~Kiddon, J.~Konecny, S.~Mazzocchi, H.~B. McMahan \emph{et~al.}, ``Towards
  federated learning at scale: System design,'' \emph{arXiv preprint
  arXiv:1902.01046}, 2019.

\end{thebibliography}
\end{document}